\documentclass[apj]{emulateapj}
\usepackage{natbib}

\newcommand{\msun}{{\rm M}_{\sun}}

\begin{document}
\shorttitle{Modeling the hard states of three black hole candidates}
\shortauthors{Zhang, Yuan, \& Chaty}

\title{Modeling the hard states of three black hole candidates}

\author{Hui Zhang\altaffilmark{1,}\altaffilmark{2}, Feng
Yuan\altaffilmark{1},and Sylvain Chaty\altaffilmark{3}}

\altaffiltext{1}{Key Laboratory for Research in Galaxies and
Cosmology, Shanghai Astronomical Observatory, Chinese Academy of
Sciences, 80 Nandan Road, Shanghai 200030, China; fyuan@shao.ac.cn}
\altaffiltext{2}{Graduate School of the Chinese Academy of Sciences,
Beijing 100039, China} \altaffiltext{3}{AIM - Astrophysique
Interactions Multi-\'echelles (UMR 7158 CEA/CNRS/Universit\'e Paris
7 Denis Diderot), CEA Saclay, DSM/IRFU/Service d'Astrophysique,
B\^at. 709, L'Orme des Merisiers, FR-91 191 Gif-sur-Yvette Cedex,
France}

\begin{abstract}
Simultaneous multiwavelength observations were recently performed
for three black hole candidates --- SWIFT J1753.5-0127, GRO J1655-40
and XTE J1720-318. In this paper, we test the accretion-jet model
originally proposed for XTE J1118+480 by investigating the hard
state of these three sources using this model. The accretion flow in
the model is composed of an inner hot accretion flow and an outer
truncated thin disk. We find that the model satisfactorily explains
the spectrum ranging from radio to X-rays, with the radio and X-ray
spectra dominated by the synchrotron and thermal Comptonization
emissions in the jet and the hot accretion flow, respectively, while
the infrared and optical being the sum of the emissions from the
jet, hot accretion flow, and the truncated thin disk. Similar to the
case of XTE J1118+480, the model can also explain, although only
qualitatively in some cases, the observed timing features including
quasi-periodic oscillation, and positive and negative time lags
between the optical and X-ray emissions detected in SWIFT
J1753.5-0127. The origin of the ejection events detected in XTE
J1720-318 is also briefly discussed.

\end{abstract}

\keywords{ accretion, accretion disks --- black hole physics---ISM:
  jets and outflows---stars: individual (SWIFT J1753.5-0127, GRO
 J1655-40, XTE J1720-318)}

\section{INTRODUCTION}

More than 40 black hole X-ray binaries or candidate black-hole X-ray
binaries have been found so far
\citep{MR06} Because they are closer to us and the black hole mass
is much smaller compared to active galactic nuclei (AGNs), they
supply us with some unique information that is very valuable to
understand the physics of accretion onto compact objects,
particularly black holes. These information include simultaneous
multiwaveband spectral energy distribution (SED), the state
transition (see below), timing features such as quasi-periodic
oscillation (QPO), and time lags between different wavebands. For
complete reviews on observations and theory of black hole X-ray
binaries, the readers can refer to McClintock \& Remillard (2006),
Fender (2006), Zdziarski \& Gierlinski (2004), and Done, Gierlinski
\& Kubota (2007). Here we only briefly introduce some of the most
relevant background.

Black hole X-ray binaries are usually thought to come into five
different states, namely quiescent, low/hard (LHS), high/soft (HSS),
intermediate, and very high (or steep power-law) states. We focus on
LHS in this paper. The quiescent state is generally assumed to be
the low-luminosity version of the LHS. The HSS state, which is
relatively well understood, is described by a standard thin disk
sandwiched by a corona. The steep power-law state or intermediate
state is still poorly understood, although some hard efforts have
been made (e.g., Kubota \& Done 2004; Done \& Kubota 2006).

One of the key mysteries of LHS has been the origin of the hard
X-ray emission, since the standard thin disk is too cool to emit
hard X-rays. One traditional explanation is from a hot corona above
and below a standard thin disk. The disk corona is hot due to
heating by reconnection of the magnetic field emerged from the
underlying disk due to Parker instability, like in the case of solar
corona (e.g., Liang \& Price 1977; Galeev, Rosner, \& Vaiana 1979).
If this is the model of the LHS (e.g., Haardt \& Maraschi 1993), we
need to answer why the hard X-ray emission in HSS is much weaker
even though the same process should also be happening in this case.
Unfortunately, while the scenario of this model is very attractive,
our current poor knowledge of magnetic reconnection hampers us from
a deeper quantitative study. The only MHD numerical simulation of
thin disk based on shearing-box approximation with self-consistent
turbulent dissipation, radiative transfer, and radiative cooling
tentatively indicates that the corona may be too weak and
temperature too low to produce the observed hard X-ray emission
(Hirose, Krolik \& Stone 2006). In addition to the disk-corona
model, another type of model for LHS is the pure jet model. In its
early version it was proposed that both radio and X-ray emissions
originate from the synchrotron emission of the jet (Markoff, Falcke
\& Fender 2001). Because the observed shape of the high-energy
cutoff cannot be easily fit by the synchrotron emission (Zdziarski
et al. 2003), the jet model evolved subsequently and now the X-ray
emission is proposed to be produced by the synchrotron-self-Compton
process in the ``jet base'' (Markoff, Nowak \& Wilms 2005). This
model has been applied to the LHS of other sources in addition to
XTE J1118+480 (Markoff, Falcke \& Fender 2001), including GRO
J1655-40, which will be studied in this paper (Migliari et al.
2007). "However, Maccarone (2005) argue that the X-rays were
unlikely to be formed in jets by the comparison of the properties of
accreting black holes and accretion neutron stars. Maitra et al.
(2009) and Malzac, Belmont \& Fabian (2009) show that the revised
jet model by Markoff et al. (2005) also possesses pair-production
problem for bright sources."

In this paper, we focus on the hot accretion flow model. In this
model, the X-ray emission is produced by a hot accretion flow via
thermal Comptonization process. The most well-known hot accretion
flow model is the advection-dominated accretion flow (ADAF; Ichimaru
1977; Rees et al. 1982; Narayan \& Yi 1994, 1995; Abramowicz et al.
1995). The ADAF model, or more generally hot accretion flow model,
has drawn intensive theoretical interests and has been successfully
applied to many low-luminosity sources since its discovery. It is
the only dynamically based accretion model to be able to produce
hard X-ray emissions (see Narayan \& McClintock 2008 and Yuan 2007
for recent reviews). Esin, McClintock \& Narayan (1997) applied the
ADAF model in detail to Nova Muscae 1991 and successfully modeled
its low-hard and quiescent states. In their model a standard thin
disk is truncated at a ``transition'' radius and replaced by an
inner ADAF. The best evidence for the truncation of thin disk is
perhaps the UV spectrum of XTE J1118+480 (Esin et al. 2001; Frontera
et al. 2001; Chaty et al. 2003; Yuan, Cui \& Narayan 2005, hereafter
YCN05. See Ho 2002 and Yuan 2007 for review of evidences for the
truncated disk in low-luminosity AGNs), which can only be fitted by
a truncated thin disk. However, this model significantly
underestimates the radio and infrared spectra (Esin et al. 2001).
This is not surprising since we now know that LHS is usually
associated with jets (Fender 2006) while they are not included in
Esin et al. (2001).

YCN05 extended the work of Esin et al. (1997; 2001) by including a
jet and applied this accretion-jet model (see Section 2 for the
description of the model) to XTE J1118+480 again because this source
has so far the best simultaneous multiwaveband data, including SED
and various kinds of timing features that supply good constraints to
theoretical models. In the interpretation of YCN05, the X-rays come
from the accretion flow (ADAF), while the radio and most of the IR
radiation come from the jet. The model not only successfully
explains the SED, but also the timing features including QPOs and
the otherwise puzzling ``negative'' and ``positive'' time lags
between optical/UV and X-rays (see also similar work by Malzac et
al. 2004). The QPO interpretation requires the existence of the
geometrically thick ADAF bounded at a certain radius while the
interpretation of time lags requires that both the jet and the ADAF
contribute to the IR and optical radiation, both of which are
natural components or results of the accretion-jet model.

An additional constraint comes from the radio-X-ray correlation
found in black hole X-ray binaries, and more generally, in black
hole sources (Gallo et al. 2003; Corbel et al. 2003; Merloni et al.
2003; Falcke, K\"ording, \& Markoff 2004; G\"{u}ltekin et al. 2009;
but see Xue \& Cui 2007 for a different point of view). This
correlation is well explained by the accretion-jet model (Yuan \&
Cui 2005). Moreover, Yuan \& Cui (2005) made two predictions. One is
that when the X-ray luminosity of the system is lower than $L_{\rm
crit} \sim 10^{-6}L_{\rm Edd}$, the correlation should steepen, with
the correlation index changing from $\sim 0.6$ to $\sim 1.23$.
Another prediction is that below $L_{\rm crit}$ the X-ray emission
of the system should be dominated by the jet. Both predictions have
recently obtained strong supports in both observational and
theoretical aspects (see review by Yuan et al. 2009c). Especially,
Yuan et al. (2009c) combined the radio and X-ray luminosity of 22
AGNs satisfying $L_{\rm x} \la L_{\rm crit}$ and found that the
correlation index is indeed $\sim 1.22$, which is in excellent
agreement with the prediction of Yuan \& Cui (2005).

The coupled ADAF-jet model has only been applied to two sources so
far, XTE J1118+480 (YCN05) and XTE J1550-564 (Yuan et al. 2007).
Thus, it is necessary to test the model on a larger sample. Recently
some new black hole candidates have been found and some of them have
very good simultaneous multiwavelength data. In this paper, we first
overview the accretion-jet model in Section 2, we then study the LHS
of three such sources, namely SWIFT J1753.5-0127, GRO J1655-40, and
XTE J1720-318, by using the main observational constraints including
SED, QPO, and time lags between different bands in Section 3. In
Section 4, we summarize our main results.

\section{The accretion-jet model}

We briefly describe the ADAF-jet model here. The readers can refer
to YCN05 for additional details. In this model, the accretion flow
at large radii is described by a standard thin disk. It makes a
transition at $R_{\rm tr}$ and becomes a hot accretion flow (ADAF).
Both observational (Yuan \& Narayan 2004) and direct theoretical
(``evaporation'' mechanism: Liu et al. 1999; R\'o\.za\'nska \&
Czerny 2000; ``turbulent diffusion'' mechanism: Manmoto \& Kato
2000) studies indicate that the value of $R_{\rm tr}$ increases with
the decrease of accretion rate, although their quantitative results
are different. Numerical simulations of hot accretion flow have
shown that the flow is convectively unstable, irrespective of
whether the radiation is weak (e.g., Igumenshchev \& Abramowicz
1999; Stone, Pringle \& Begelman 1999; 2001) or strong (Yuan \& Bu
2010). As a result, the accretion rate decreases inward because the
gas circulates in the convective eddies. To mimic this effect, we
parameterize the accretion rate with a parameter $s$, $\dot{M} =
\dot{M}_{\rm out}(R/R_{\rm tr})^s$, where $\dot{M}_{\rm out}$ is the
accretion rate at the outer boundary of the hot accretion flow
$R_{\rm tr}$ and $s$ describes the strength of the outflow. For the
viscous parameter $\alpha$ and magnetic parameter $\beta$ (defined
as the ratio of gas pressure to the sum of gas and magnetic
pressure), we set $\alpha= 0.3$ and $\beta = 0.9$ as suggested by
MHD numerical simulations (Hawley \& Krolik 2001). Hence they are
not free parameters in our model. Another parameter is $\delta$, the
fraction of the turbulent dissipation that directly heats the
electrons. The values of $\delta$ and $s$ are well constrained in
the case of our Galactic center supermassive black hole, Sgr A*
(Yuan et al. 2003), which are $\delta= 0.5$ and $s = 0.3$. However,
we want to emphasize that large uncertainties exist in their values.
For example, some numerical simulations indicate moderately smaller
$\delta$ ($\sim (T_e/T_i)^{1/2}$: Sharma et al. 2007) and larger $s$
($0.5 \la s \la 1$: Stone \& Pringle 2001). We therefore also try
other values. The radiative processes in hot accretion flow include
synchrotron, bremsstrahlung, and their Comptonization. We first
solve for the global dynamical solution of the hot accretion flow to
obtain the physical quantities such as density, ion and electron
temperature, and magnetic field. We then calculate the radiative
transfer to obtain the emitted spectrum. The details of the
calculation can be found in Yuan et al. (2003). Specifically, the
X-ray radiation is typically dominated by thermal Comptonization for
most of the accretion rates of interest.

The model of jet radiation is based on the internal shock scenario,
which has also been widely adopted in the study of gamma-ray bursts.
A small fraction of the material in the accretion flow, described by
$\dot{M}_{\rm jet}$, is transferred into the vertical direction and
forms a jet. The jet is assumed to have a conical geometry with a
half-open angle $\phi$ and a bulk Lorentz factor $\Gamma_{\rm jet}$.
We fix those values to be $\phi= 0.1$ and $\Gamma_{\rm jet} = 1.2$,
thus they are not free parameters. A larger $\phi$ will decrease the
total flux, but this impact can be absorbed by the accretion rate.
$\Gamma_{\rm jet}$ is well within the range obtained by combining
observations and numerical simulations: $\Gamma_{\rm jet} \le 1.67$
(Gallo et al. 2003). A small fraction of the electrons in the jet
are accelerated by the internal shocks and form a power-law energy
distribution with index $p$. The energy density of accelerated
electrons and amplified magnetic field in the shock front is
determined by two free parameters, $\epsilon_e$ and $\epsilon_B$,
defined as the ratio of the electron energy and magnetic energy to
the shock energy, respectively. Only synchrotron emission is
considered since Compton scattering is not important in our model
(Markoff, Falcke \& Fender 2001; Wu, Yuan \& Cao 2007). The radio
and X-ray radiation comes from the optically thick and optically
thin synchrotron emission of the accelerated electrons,
respectively.

The main advantage of the accretion-jet model is that it has strong
dynamical basis, such as the thin disk, hot accretion flow, and the
transition between them. We would like to point out, however, that
large uncertainties exist in our model since many simplifications
and assumptions have to be adopted in current models of accretion
flow and jet. First, accretion flows are intrinsically
three-dimensional but we so far can only get global solution under
one-dimensional approximation. Second, the exact value of the
viscous parameter $\alpha$ is still unknown and it must be a
function of radius. It very likely originates from the stress of the
magnetic field amplified by Magneto-Rotational Instability (MRI)
(Balbus \& Hawley 1998). MHD numerical simulation can produce some
values but they depend on, e.g., initial magnetic field
configuration which is unclear. Third, general relativity effect,
which is important in the inner region of the accretion flow and the
jet, is usually also neglected. Fourth, when calculating
Comptonization, almost all current work consider only local
scattering while the global effect is neglected (Yuan et al. 2009b;
Xie et al. 2010). Fifth, the jet model is even worse; the formation
and collimation mechanisms of jet are still not solved, even the
ingredient of the jet is debated. In spite of the above caveats, all
the above effects are of higher order and we believe that our
current models can present valuable information when we compare
their prediction with observations. On the other hand, given these
model uncertainties, we think that it is physically not appropriate
to do fine statistical analysis such as $\chi^2$ analysis when
comparing the theoretical prediction with data as adopted in some
works (e.g., Migliari et al. 2007). When judging the goodness of the
modeling, we have to be satisfied with comparison by eye. But in
Section 3.4 we do show the effects of changing the main model
parameters, which are $\dot{M}, \delta, s$, and $\dot{M}_{\rm jet}$.

\section{Modeling the three sources}

\subsection{SWIFT J1753.5-0127}

\subsubsection{Spectrum}

SWIFT J1753.5-0127 was discovered on 2005 May 30 when it underwent
an X-ray outburst. According to its spectrum and timing, this source
is regarded as a good black hole candidate (e.g., Miller et al.
2006; Ramadevi \& Seetha 2007; Zhang et al. 2007; Cadolle Bel et al.
2007). Simultaneous multiwavelength observations were performed with
{\it International Gamma-Ray Astrophysics Laboratory (INTEGRAL)},
{\it RXTE}, {New Technology Telescope} ({NTT}), {Rapid Eye Mount
Telescope}, and {Very Large Array (VLA)} on 2005 August 10-12, and a
good multiwaveband spectrum was obtained (Cadolle Bel et al. 2007).
The data obtained on August 11 are shown in Figure 1, which are
taken from Cadolle Bel et al. (2007).

It is still difficult to determine the accurate parameters, such as
mass, distance and inclination angle of this source, so following
Cadolle Bel et al. (2007), we set the distance of the source $d=6$
kpc, the mass of the black hole $M=9\msun$, and the inclination
angle of the source to be 63$^\circ$. Changing these parameters will
have different effects on our modeling results. A larger value of
the inclination angle will make the emission from the jet weaker,
but most of this effect can be absorbed by the mass-loss rate in the
jet. A larger $M$ or smaller $d$ will make the accretion rate
smaller thus the predicted spectrum softer, if all other model
parameters are kept fixed. We then calculate the emitted spectrum
from the ADAF and jet as described above. The top panel of Figure 1
shows our best modeling results. The values of parameters adopted
are $\dot{M}=0.1\dot{M}_{\rm Edd}\equiv L_{\rm Edd}/c^2$, $s=0.3$,
$\delta=0.5$, $R_{\rm tr}=250\,R_s$, $\dot{M}_{\rm jet}=8\times
10^{-5}\dot{M}_{\rm Edd}$, $\epsilon_e=0.04$, $\epsilon_B=0.02$, and
$p=2.1$. We can see from the figure that, similar to XTE J1118+480
(YCN05) and XTE J1550-564 (Yuan et al. 2007), the radio and X-ray
emissions are dominated by the jet and the ADAF, respectively, while
the infrared (IR) and optical are the sum of the emissions from the
jet, ADAF, and the truncated thin disk. Given that both accretion
flow and jet models have large uncertainties and some approximations
have to be adopted, as we have stated in Section 2, we think it is
unreasonable to do elaborate statistical analysis as was done in
Migliari et al. (2007). To show the ``goodness'' of the fitting we
present the ratio of the observed flux and the theoretical
prediction in the bottom panel of Figure 1 (and Figures 2 and 3 as
well). For the observed flux, we simply adopt the middle value but
neglect their errorbar. We feel the modeling results are
satisfactory.

\begin{figure}
\epsscale{1.0} \plotone{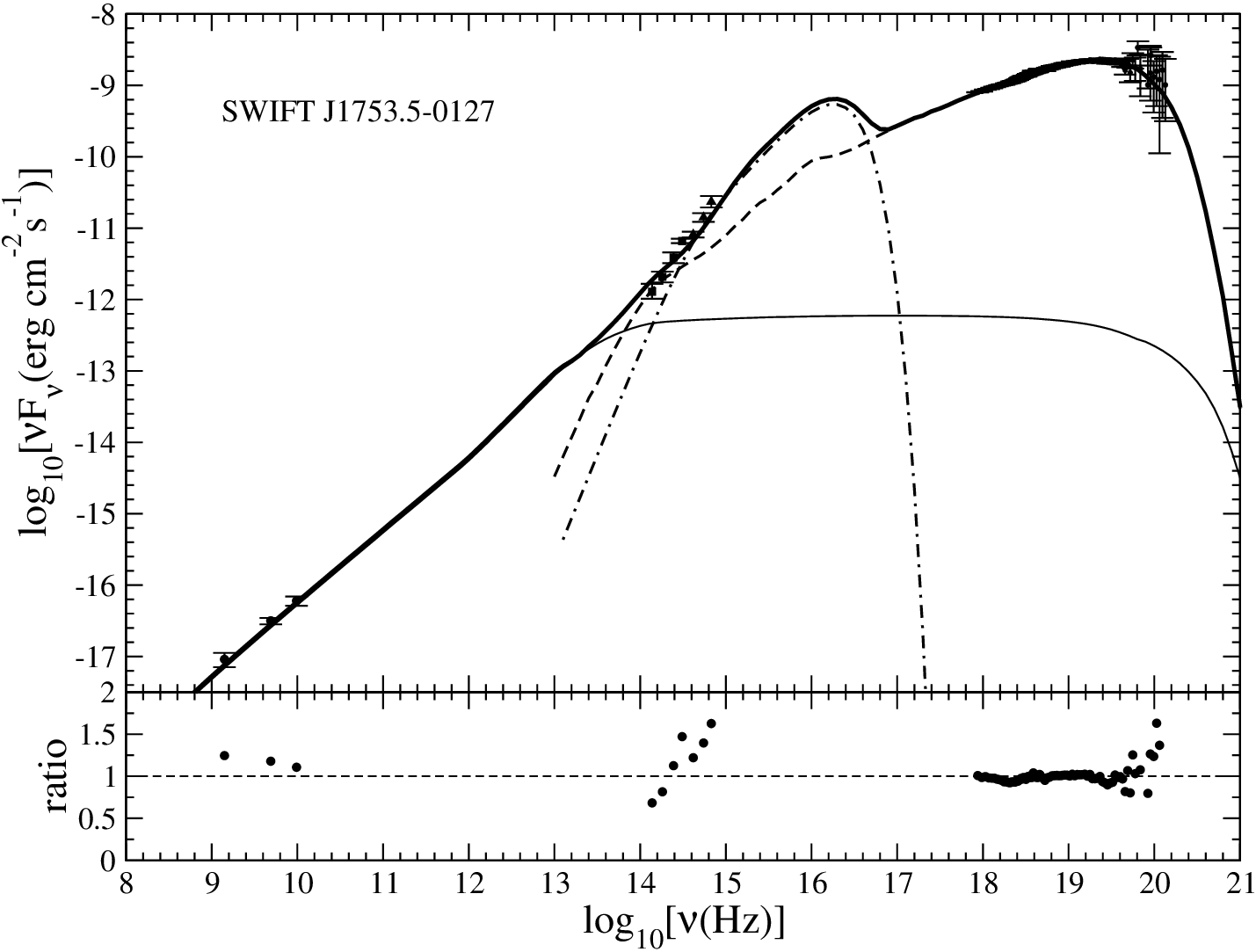} \label{fig:j1753} \vspace{0.5cm}
\caption{SED of SWIFT J1753.5-0127. {\it Top:} the dash-dotted,
dashed, and thin solid lines show the emissions of the truncated
thin disk, the hot accretion flow, and the jet, respectively. The
thick solid line shows their sum. {\it Bottom:} ratio of the
observed flux and the theoretical value.}
\end{figure}

\subsubsection{Timing features: QPO and the time lags between
optical and X-ray emissions}

A low-frequency QPO was clearly detected and its frequency is 0.241
Hz on August 11 (Cadolle Bel et al. 2007). The QPO frequency is
found to correlate with the X-ray flux almost linearly (Zhang et al.
2007; Ramadevi \& Seetha 2007). Such a correlation seems to be
common in black hole X-ray binaries (e.g., Cui et al. 1999 for XTE
J1550-564).

Many models of QPO have been proposed. Especially, for the accretion
configuration of our model, namely an inner hot accretion flow plus
an outer truncated thin disk, Giannios \& Spruit (2004; see also
Rezzolla et al. 2003) suggested that the QPO can be excited by the
interaction of ADAF and the thin disk and resulted from the basic
{\it p}-mode oscillations of the inner ADAF, with a frequency near
the Keplerian frequency at $R_{\rm tr}$. The Keplerian frequency at
$R_{tr}=250\,R_s$ is 0.32 Hz, which is close to the observed QPO
frequency of 0.24Hz. During the outburst, the decrease of the
accretion rate results in the decrease of the X-ray flux. Since the
value of $R_{\rm tr}$ increases (thus the Keplerian frequency at
$R_{\rm tr}$ decreases) in this process, as introduced in Section 2,
this qualitatively explains the detected correlation between the QPO
frequency and the X-ray flux.

Durant et al. (2008) conducted simultaneous optical and X-ray
observations and did cross-correlation analysis. They found that the
optical emission {\em decreases} a few seconds {\em before} the
X-rays {\em increase}. In addition to this ``negative time lag'' or
``precognition dip'', the cross-correlation function also shows a
weaker ``positive lag'', which means that the optical emission lags
the X-rays by several seconds. These positive and negative lags are
similar to XTE J1118+480, but the timescales of these two lags are
markedly different. For XTE J1118+480, the precognition dip or
negative lag is weaker while the positive lag is stronger (Kanbach
et al. 2001).

The two time lags in XTE J1118+480 have been explained by the
accretion-jet model in YCN05. The same mechanism works for SWIFT
J1753.5-0127. From Figure 1, we can see that both  ADAF and the jet
contribute to the optical emission. When there is a perturbation
such as a sudden increase of accretion rate, it will first propagate
in ADAF, then into the jet. The increase of accretion rate and
subsequently the increase of mass-loss rate in the jet will result
in an increase of X-ray emission from ADAF and subsequently an
increase of optical emission from the jet. This is the reason for
the positive lag. On the other hand, ADAF emits both optical and
X-ray emissions. The optical emission originates from the
self-absorbed synchrotron emission, which depends on the profiles of
$T_e$ and optical depth. For the specific parameters of the hot
accretion flow in SWIFT J1753.5-0127, we find that an {\em increase}
of $\dot{M}$ results in a {\em decrease} of the optical flux. Since
optical emission comes from $\sim 40\,R_s$ while the X-rays from
$\sim 10\,R_s$, an increase of $\dot{M}$ first results in a decrease
of optical flux then an increase of X-ray flux after the accretion
timescale at $\sim 40\, R_s$. This could be the reason for the
negative lag.

The next question is why the intensity of the two time lags are
opposite. This is because of the differences in the model
parameters. The value of $\dot{M}$ in SWIFT J1753.5-0127 is higher
than that in XTE J1118+480 ($\dot{M}=0.05\dot{M}_{\rm Edd}$), while
$\dot{M}_{\rm jet}$ is lower than XTE J1118+480 ($\dot{M}_{\rm
jet}=2.5\times 10^{-4}\dot{M}_{\rm Edd}$). Comparing Figure 1 in
this paper with Figure 2 in YCN05, we can see that this results in a
significantly smaller relative contribution of the jet compared to
ADAF in SWIFT J1753.5-0127 than in XTE J1118+480. This is why the
intensities of the ``positive'' and ``negative'' correlation are
opposite between the two sources.

Quantitatively, however, we can only explain the positive time lag.
The optical emission in the jet mainly comes from $d \sim 7000$
$R_s$. The X-ray emission from ADAF originates from $\la 10R_s$. So
the timescale of the positive lag is determined by $d/c\sim 1.5$s,
which roughly agrees with the observation. On the other hand, the
optical emission by ADAF mainly comes from $\sim 40 R_s$. Thus the
timescale of the negative lag is determined by the accretion
timescale there, which is $\sim 0.2$s. This is much shorter than the
detected timescale of several seconds. We speculate that the
discrepancy may be due to the following reasons. For technical
reasons, when solving for the global solution of ADAF, we have to
set the angular velocity at $R_{\rm tr}$ to be substantially
sub-Keplerian, although it should be super-Keplerian there. This
makes the radial velocity of the accretion flow much larger than it
should be and thus the accretion time scale is much shorter. Another
possible reason is that we choose the viscous parameter $\alpha=0.3$
in our calculations, but the actual value may be significantly
smaller .

\subsection{GRO J1655-40}

GRO J1655-40 entered a new outburst in 2005 February after seven
years of quiescence and many observations have been conducted then
(e.g., Shaposhnikov et al. 2007; Caballero-Garc\'{\i}a et al. 2007;
Joinet, Kalemci, \& Senziani 2008) especially a simultaneous
multiwavelength campaign by {\it RXTE}, {\it INTEGRAL}, {\it SMART},
{VLA}, and {\it Spitzer} (Migliari et al. 2007). We make use of the
multiwavelength data obtained on September 24 when the source was in
the ``standard'' LHS, as shown in Figure 2. The details of
observations and  data analysis can be found in Migliari et al.
(2007). A QPO with frequency of 0.3 Hz was also found that day
(Migliari et al. 2007).

\begin{figure}
\epsscale{1.0} \plotone{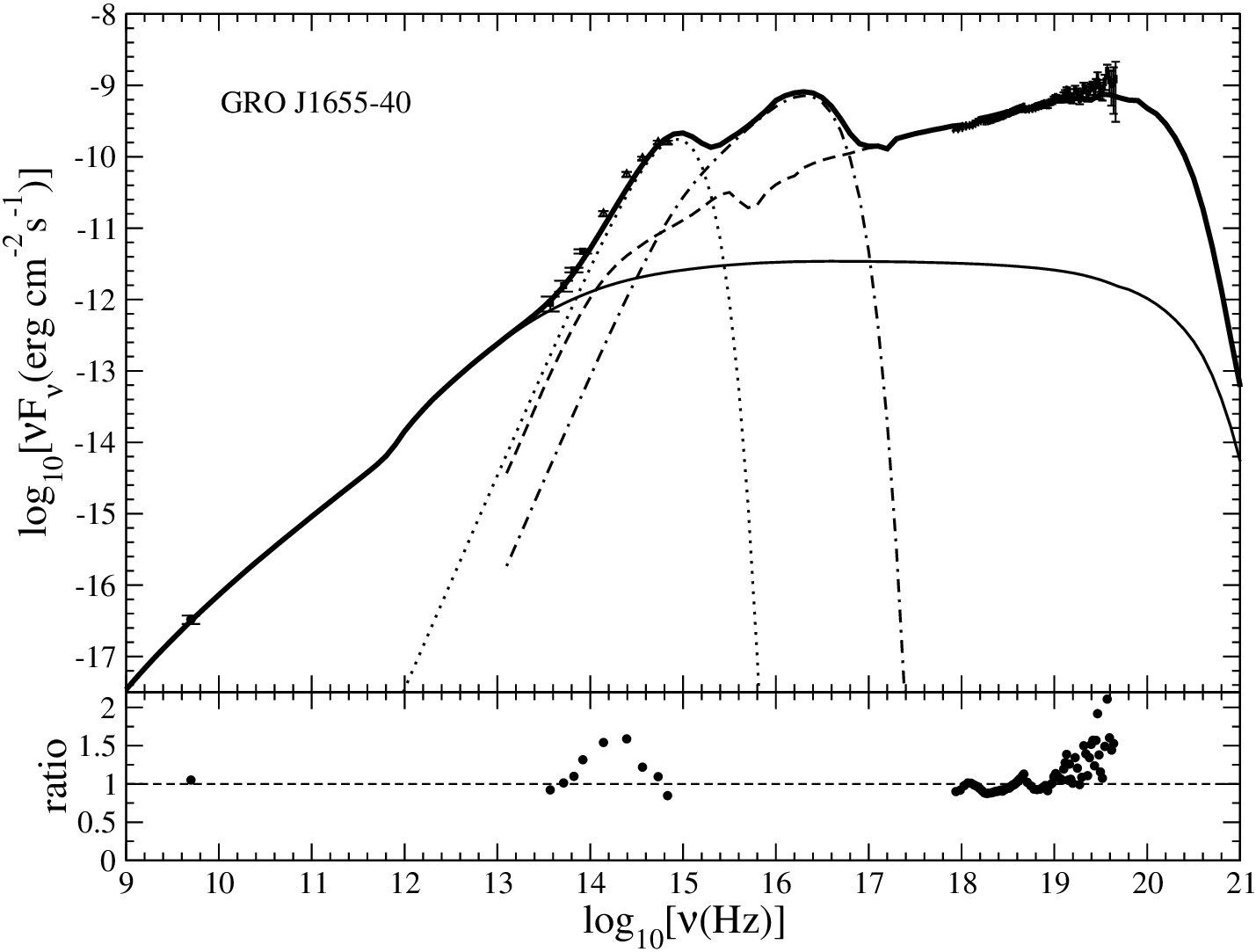} \label{fig2} \caption{SED of GRO
J1655-40. {\it Top:} the dash-dotted, dashed, and thin solid lines
show the emissions of the truncated thin disk, the hot accretion
flow, and the jet, respectively. The dotted line shows the blackbody
emission of the companion star. The thick solid line shows their
sum. {\it Bottom: ratio of the observed flux and the theoretical
value.}}
\end{figure}

\begin{figure}
\epsscale{1.0} \plotone{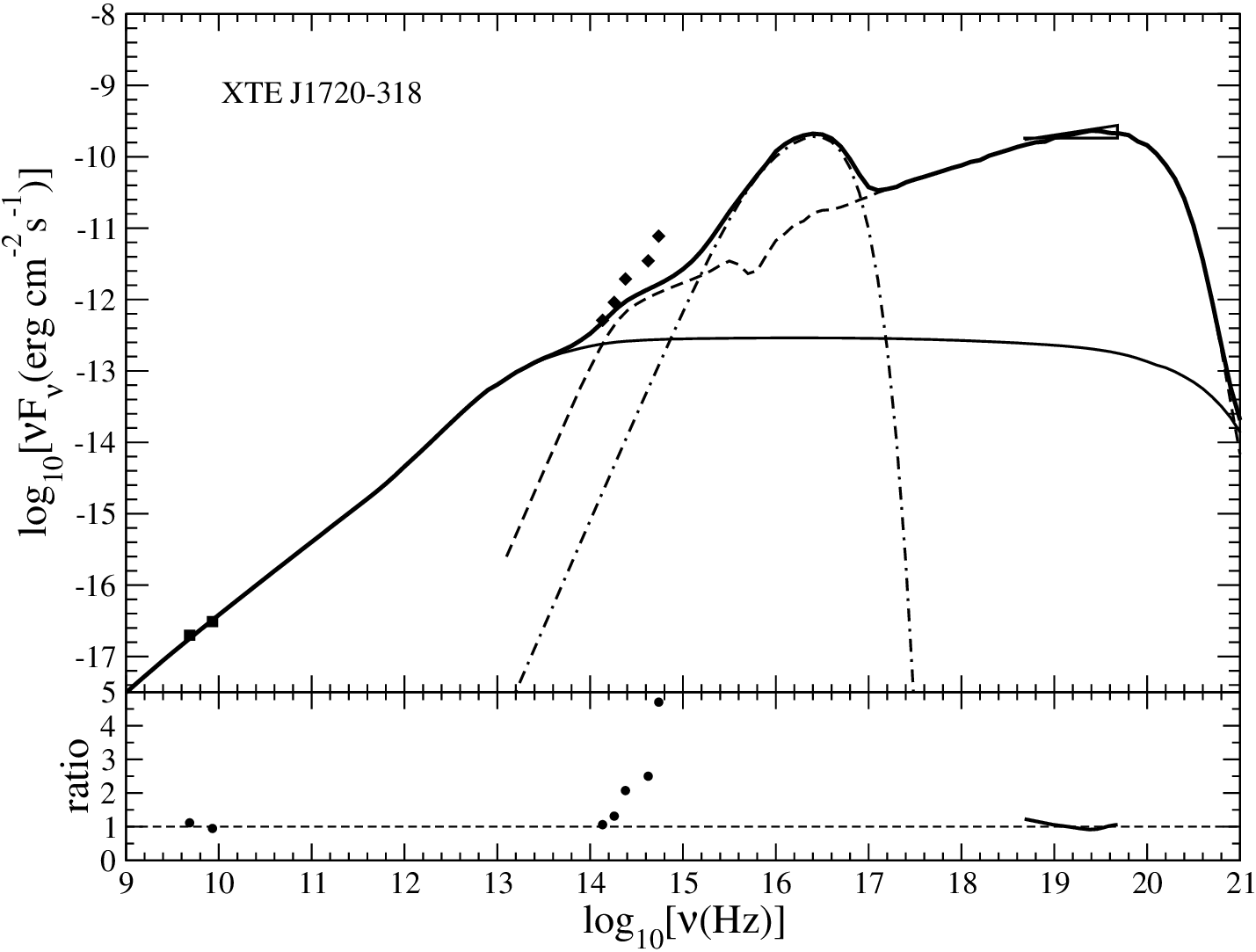} \vspace{0.0cm} \caption{SED of XTE
J1720-318. {\it Top:} the dash-dotted, dashed, and thin solid lines
show the emissions of the truncated thin disk, the hot accretion
flow, and the jet, respectively. The thick solid line shows their
sum. Note the contribution from the companion star has not been
subtracted in the data. {\it Bottom: ratio of the observed flux and
the theoretical value.}}
\end{figure}

To model this source, we adopt the mass of the black hole
$M=6.3\msun$ (Greene et al. 2001), and the distance $D=3.2$ kpc
(Hjellming \& Rupen 1995). The inclination angle of the accretion
disk is $69^{\circ}.50\pm 0^{\circ}.08$(Orosz \& Bailyn 1997), while
Greene obtained $70^{\circ}.2\pm 1^{\circ}.9$ (Greene et al. 2001).
So we set the inclination angle of this source to be $70^{\circ}$ in
this paper. Other values of mass and inclination angle are reported
by Beer \& Podsiadlowski (2002), but their values do not deviate too
much from our values adopted here and will not impact our final
result seriously. The inclination angle of the jet is
$85^{\circ}\pm2^{\circ}$ (Hjellming \& Rupen 1995), which means that
the jet is not perpendicular to the disk, and here we set the
inclination angle of the jet to be $85^{\circ}$ for jet modeling.

Figure 2 shows the modeling result of the spectrum. In addition to
the accretion flow and jet, also shown in the figure is the emission
from the possible companion star, since the impact of the companion
star has not been subtracted. The model parameters are
$\dot{M}=0.06\dot{M}_{\rm Edd}$, $R_{\rm tr}=200\,R_s$,
$s=\delta=0.3$, $\dot{M}_{\rm jet}=2\times 10^{-4}\dot{M}_{\rm
Edd}$, $\epsilon_e=0.06$, $\epsilon_B=0.02$, and $p=2.1$. Again, the
radio and X-ray emissions are dominated by the jet and ADAF,
respectively. The IR and optical emissions are the sum of the
emissions of the jet, ADAF, truncated thin disk, and the companion
star.

The Keplerian frequency at $R_{\rm tr}=200\,R_s$ is $\sim 0.45$Hz,
which is roughly consistent with the frequency of the detected QPO.

\subsection{XTE J1720-318}
\subsubsection{Spectrum}

XTE J1720-318 is a black hole candidate discovered in 2003. Nearly
simultaneous multiwavelength observations were conducted during its
outburst in 2003 with VLA in radio (April 26; Brocksopp et al.
2005), NTT and SOFI in optical and IR (April 24 and 27; Chaty \&
Bessolaz 2006), and {\it INTEGRAL} in X-ray (April 6-22; Cadolle Bel
et al. 2004). The spectrum is shown in Figure 3. Note that the
contribution from the companion star has not been subtracted.

Following Cadolle Bel et al. (2004) and Chaty \& Bessolaz (2006), we
assume the mass of the black hole, distance, and inclination angle
to be $M=5\msun$, $d=8$ kpc and $i=60^{\circ}$, respectively. Figure
3 shows the modeling results. The model parameters are
$\dot{M}=0.08\dot{M}_{\rm Edd}$, $s=\delta=0.5$, $R_{\rm tr}=150\,
R_s$, $\dot{M}_{\rm jet}=6\times10^{-5}\dot{M}_{\rm Edd}$,
$\epsilon_e=0.06$, $\epsilon_B=0.08$, and $p=2.1$. The modeling
results are similar to the other two sources, i.e. the radio and
X-ray emissions are dominated by the jet and ADAF, respectively.

\subsubsection{Ejection events}

An interesting result from radio observation is that two ejection
events took place, possibly associated with the state transition
from hard to soft (Brocksopp et al. 2005). The spectral analysis
indicates that the radio-emitting material expands and becomes
optically thin from optically thick. Actually such kind of episodic
ejection of material seems to be common during the transition from
hard to soft states (see review in Fender, Belloni \& Gallo 2004).
The ejection is regarded as a ``type-II'' jet and its features are
reviewed in detail in Fender \& Belloni (2004). An interesting
question is then whether we can explain these ejection events in the
framework of the accretion-jet model?

The formation of continuous jet or ``type-I'' jet has been the topic
of intensive study for many years and we now have relatively good
understanding of it. Briefly, the formation of this kind of jet is
due to the extraction of the rotation energy of the spinning black
hole (Blandford \& Znajek 1977)  or the accretion disk (Blandford \&
Payne 1982) by a large-scale {\em open} magnetic field. In contrast,
we have little knowledge of the formation of episodic jets. Very
recently, by analogy with the ``coronal mass ejection'' theory in
solar physics, Yuan et al. (2009a) proposed that the formation of
episodic jets is due to the extraction of rotation energy of the
accretion flow by the {\em closed} magnetic field amplified in the
disk and emerged into the corona. During the transition from LHS to
HSS, the hot accretion flow collapses and forms a thin disk due to
the strong cooling. In this process, a large amount of magnetic flux
emerges out of the accretion flow in a very short timescale and some
coronal material is pushed outward by the strong magnetic pressure
force. For details, the readers can refer to Yuan et al. (2009a).

\begin{figure}
\epsscale{1.0} \plotone{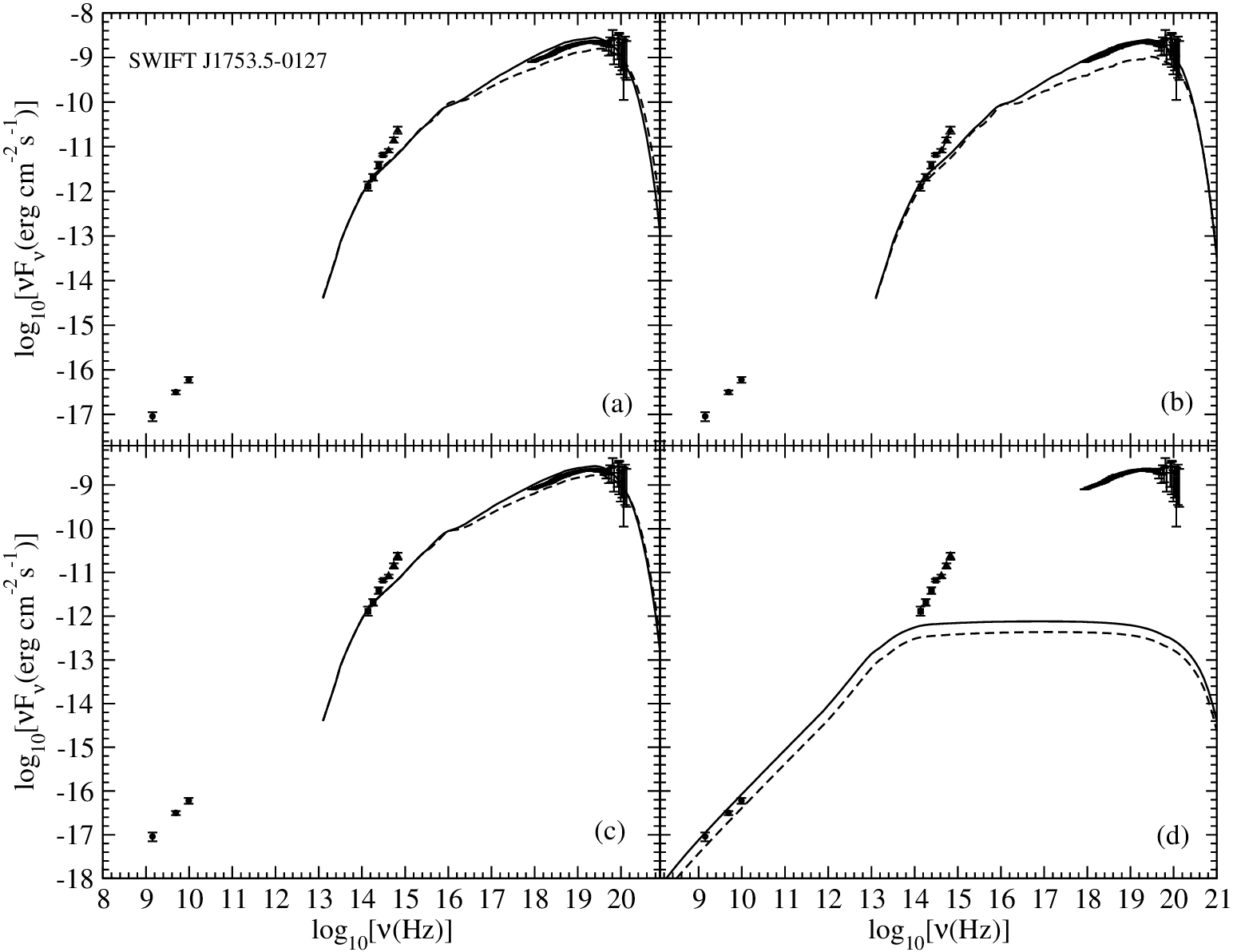} \vspace{0.0cm} \caption{Effects of
changing main model parameters for SWIFT J1753.5-0127. Each panel
shows the larger and smaller parameters than the ``best'' one. Panel
(a): $\dot{M}=0.105\dot{M}_{\rm Edd}$ (solid line) and
$\dot{M}=0.09\dot{M}_{\rm Edd}$ (dashed line). Panel (b):
$\delta=0.6$ (solid line) and $\delta=0.3$ (dashed line). Panel (c):
$s=0.2$ (solid line) and $s=0.5$ (dashed line). Panel (d):
$\dot{M}_{\rm jet}=10^{-4}\dot{M}_{\rm Edd}$ (solid line) and
$\dot{M}_{\rm jet}=6\times10^{-5}\dot{M}_{\rm Edd}$ (dashed line).}
\end{figure}

\begin{figure}
\epsscale{1.0} \plotone{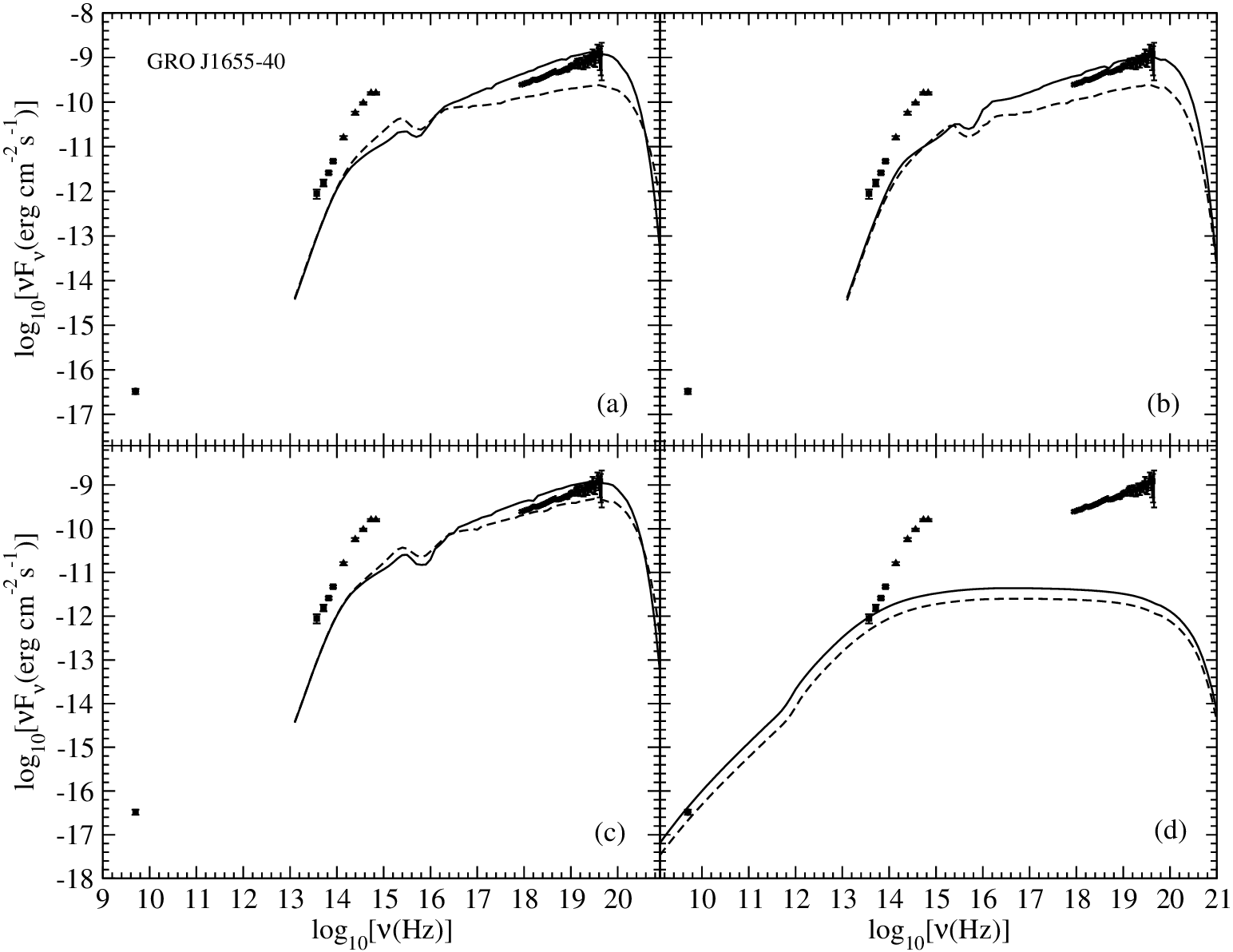} \vspace{0.0cm} \caption{Effects of
changing main model parameters for GRO J1655-40. Each panel shows
the larger and smaller parameters than the ``best'' one. Panel (a):
$\dot{M}=0.07\dot{M}_{\rm Edd}$ (solid line) and
$\dot{M}=0.05\dot{M}_{\rm Edd}$ (dashed line). Panel (b):
$\delta=0.5$ (solid line) and $\delta=0.1$ (dashed line). Panel (c):
$s=0.1$ (solid line) and $s=0.5$ (dashed line). Panel (d):
$\dot{M}_{\rm jet}=2.5\times10^{-4}\dot{M}_{\rm Edd}$ (solid line)
and $\dot{M}_{\rm jet}=1.5\times10^{-4}\dot{M}_{\rm Edd}$ (dashed
line).}
\end{figure}

\begin{figure}
\epsscale{1.0} \plotone{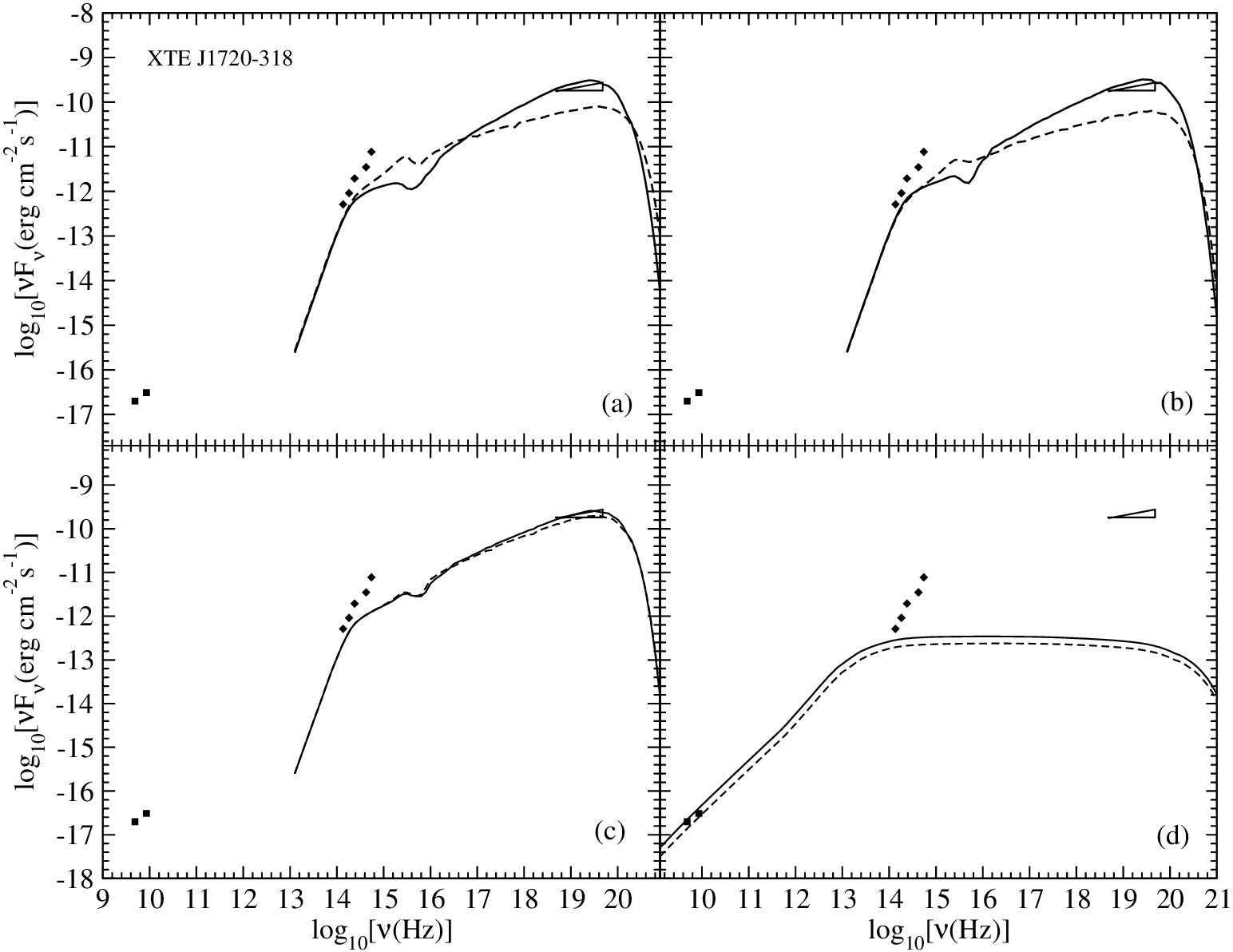} \vspace{0.0cm} \caption{Effects of
changing main model parameters for XTE J1720-318. Each panel shows
the larger and smaller parameters than the ``best'' one. Panel (a):
$\dot{M}=0.09\dot{M}_{\rm Edd}$ (solid line) and
$\dot{M}=0.07\dot{M}_{\rm Edd}$ (dashed line). Panel (b):
$\delta=0.6$ (solid line) and $\delta=0.3$ (dashed line). Panel (c):
$s=0.3$ (solid line) and $s=0.6$ (dashed line). Panel (d):
$\dot{M}_{\rm jet}=7\times10^{-5}\dot{M}_{\rm Edd}$ (solid line) and
$\dot{M}_{\rm jet}=5\times10^{-5}\dot{M}_{\rm Edd}$ (dashed line).}
\end{figure}

\subsection{The effects of changing model parameters}

As introduced in Section 2, the main free parameters of our model
are $\dot{M},\delta,s$ and $\dot{M}_{\rm jet}$. It is therefore
interesting to see how the modeling results (i.e., the predicted
spectrum) will change when these parameters are changed. Based on
the ``best'' model of each source as shown in Figures 1-3, we change
one of the above four parameters each time by a larger and smaller
value compared to the ``best'' value of the model, but keep all
other parameters the same. We calculate the emitted spectra and show
the results in Figures 4-6 for the three sources. As we expect, a
higher $\dot{M}, \delta, \dot{M}_{\rm jet}$ and a lower $s$ will
result in a higher emitted flux, a lower $\dot{M}, \delta,
\dot{M}_{\rm jet}$, and a higher $s$ will result in a lower emitted
flux. The difference of the lines in each figure gives us an idea of
the dependence or sensitivity of the modeling result on the
parameters.

\section{Summary and Discussion}

YCN05 proposed an accretion-jet model for the LHS of XTE J1118+480.
In this model the outer thin disk makes a transition at $R_{\rm tr}$
and becomes a hot accretion flow. At the innermost region some
fraction of the accretion gas is transferred into the vertical
direction and forms a jet. This model successfully explained the
spectral and timing features of XTE J1118+480 (YCN05) and XTE
J1550-564 (Yuan et al. 2007). In this paper we apply the model to
three additional sources, namely SWIFT J1753.5-0127, GRO J1655-40,
and XTE J1720-318, for which we have new simultaneous multiwaveband
spectral and timing data. Similar to our previous results, we find
that the radio and X-ray spectra are dominated by the synchrotron
emission from the jet and the thermal Comptonization in the hot
accretion flow, respectively; while the IR and optical emissions are
the sum of emissions of the jet, ADAF, and the truncated thin disk.
The QPO is explained by the oscillation of ADAF, with the QPO
frequency close to the Keplerian frequency at $R_{\rm tr}$. The
``positive'' and ``negative'' time lags between optical and X-ray
radiations can also be explained by the model, although only
qualitatively so far in the case of the ``negative lag''(Section
3.1.2).

SWIFT J1753.5-0127 and XTE J1720-318 are two of the so-called
outliers to the radio/X-ray correlation, in the sense that their
radio emission is very weak compared to their X-ray emission. The
radio luminosity of these two sources is $\sim$ 2 and 3 orders of
magnitude lower respectively than that predicted by the general
radio/X-ray correlation introduced in Section 1 and their X-ray
luminosity. This is reflected by the lower ratio of $\dot{M}_{\rm
jet}$ and $\dot{M}$ compared to that of GRO J1655-40. We speculate
that this indicates that the spin of the black holes, $a$, in these
two sources is small compared to other sources. This makes the jets
very weak since it has been shown that jet power is proportional to
$a^4$ or even $a^6$ (Tchekhovskoy, Narayan \& McKinney 2010; but see
also Fender, Gallo \& Russell 2010).

Miller et al. (2006) analyzed the high-resolution spectral data of
the LHS of Swift J1753.5-0127 and found a soft or thermal component
when the 0.5-10 keV luminosity is higher than $\sim 3\times
10^{-3}L_{\rm Edd}$. They interpret it as the emission of a standard
thin disk extending to the innermost stable circular orbit (ISCO).
This conclusion, however, is currently debated. Gierlinski, Done \&
Page (2008) analyzed the same data set of a source and showed that
the spectrum can also be fitted without the need for a thermal
component. Hiemstra et al. (2009) tested more spectral models than
Miller et al. and concluded that several models can fit the data
without the need for the thermal component. Even though the data
analysis of Miller et al. is correct, the existence of a thermal
component does not necessarily imply that the thin disk must extend
to the ISCO. This is because the component could also originate from
some cold clumps embedded in the hot accretion flow, formed due to
thermal instability when the accretion rate is higher than roughly
the critical accretion rate of an ADAF $\sim \alpha^2\dot{M}_{\rm
Edd} \equiv \alpha^2 10L_{\rm Edd}/c^2$ (Yuan 2003; Yuan et al.
2007); or from a small inner disk, formed due to the strong cooling
and condensation of the hot accretion flow (Liu et al. 2007; Taam et
al. 2008).

\acknowledgments

We thank M. Cadolle Bel and S. Migliari for providing us with the
data for SWIFT J1753.5-0127 and GRO J1655-40, respectively. We also
thank the second referee's constructive suggestions for the
presentation of the paper. This work was supported in part by the
Natural Science Foundation of China (grants 10773024, 10833002,
10821302, and 10825314), the National Basic Research Program of
China (973 Program 2009CB824800), and the CAS/SAFEA International
Partnership Program for Creative Research Teams.

\end{document}